\documentclass[sigconf]{acmart}

\usepackage{booktabs} 

\setcopyright{rightsretained}

\usepackage{balance}
\usepackage{tikz}
\usepackage{subcaption}
\usepackage{collcell}
\usepackage{colortbl}

\def\colorModel{hsb} 
\newcommand\ColCell[2]{
    \pgfmathparse{#2/#1 < 0.4?1:0}  
    \ifnum\pgfmathresult=0\relax\color{white}\fi
        \pgfmathsetmacro\compA{0}               
        \pgfmathsetmacro\compB{0}               
        \pgfmathsetmacro\compC{1-#2/#1}         
    \edef\x{\noexpand\centering\noexpand\cellcolor[\colorModel]{\compA,\compB,\compC}}\x #2
}
\newcolumntype{E}{>{\collectcell{\ColCell{500}}}m{0.4cm}<{\endcollectcell}}
\newcolumntype{F}{>{\collectcell{\ColCell{10}}}m{0.4cm}<{\endcollectcell}}

\begin{document}
\title{CNN-based Facial Affect Analysis on Mobile Devices}

\author{Charlie Hewitt}
\affiliation{\institution{University of Cambridge}}
\email{cth40@cam.ac.uk}

\author{Hatice Gunes}
\affiliation{\institution{University of Cambridge}}
\email{hatice.gunes@cl.cam.ac.uk}

\begin{abstract}
This paper focuses on the design, deployment and evaluation of Convolutional Neural Network (CNN) architectures for facial affect analysis on mobile devices. Unlike traditional CNN approaches, models deployed to mobile devices must minimise storage requirements while retaining high performance. We therefore propose three variants of established CNN architectures and comparatively evaluate them on a large, in-the-wild benchmark dataset of facial images. Our results show that the proposed architectures retain similar performance to the dataset baseline while minimising storage requirements: achieving 58\% accuracy for eight-class emotion classification and average RMSE of 0.39 for valence/arousal prediction. To demonstrate the feasibility of deploying these models for real-world applications, we implement a music recommendation interface based on predicted user affect. Although the CNN models were not trained in the context of music recommendation, our case study shows that: (i) the trained models achieve similar prediction performance to the benchmark dataset, and (ii) users tend to positively rate the song recommendations provided by the interface. Average runtime of the deployed models on an iPhone 6S equates to $\sim$45 fps, suggesting that the proposed architectures are also well suited for real-time deployment on video streams.
\end{abstract}

%
%

\ccsdesc[500]{Human-centered computing~Mobile devices}
\ccsdesc[500]{Human-centered computing~Interactive systems and tools}

\keywords{Affective computing, mobile computing, intelligent user interfaces, facial affect analysis, emotions, arousal, valence, music recommendation.}

\maketitle{}

\section{Introduction}\label{sec:int}
\subsection{Motivation and Contributions}

Affective computing has historically remained confined to laboratory settings, typically only involving small studies and with little in the way of large-scale practical application. Recent advances in deep machine learning techniques and increasing availability of large, in-the-wild datasets have led to improved performance in affect recognition tasks such as prediction of emotion, valence and arousal from facial images in real world scenarios, not just in constrained environments. The ubiquitousness of mobile devices with advanced sensors, including high quality cameras, means that the application of affective computing technologies to end-user applications is now a real possibility~\cite{surveymobileaff}.

This paper aims to explore the feasibility of obtaining state-of-the-art facial affect analysis from a captured facial image using machine learning approaches within the constrained environment of a mobile device, as well as how readily the output of these models can be used within a mobile application. To this aim, we developed the Emosic application which prompts the user to take an image of their face and predicts the displayed prominent facial affect in terms of an emotion category---neutral, happy, sad, surprised, afraid, disgusted, angry or contemptuous---as well as levels of valence (i.e., how positive/negative the displayed affect is) and arousal (i.e., how active/inactive the displayed affect is) using convolutional neural network (CNN) models. Based on the predicted user affect, the Emosic application presents a number of recommended songs to the user. Essentially our work has two primary contributions: (i) design and comparative evaluation of three CNN architectures for mobile affect analysis using the newly available AffectNet dataset~\cite{affectnet}; and (ii) demonstration of deployability of the trained models for music recommendation.

The Emosic application is intended as a case study and a proof-of-concept that emotionally intelligent user interfaces (EIUI) on mobile devices are now feasible using modern machine learning approaches and large, in-the-wild datasets. Although the CNN models were not trained in the context of music recommendation, our case study shows that: (i) the trained models overall achieve similar prediction performance to the benchmark dataset, and (ii) users positively rate the song recommendations provided by the interface.

The rest of the paper is organised as follows. A summary of related work is presented in the remainder of Sec.~\ref{sec:int}. Preprocessing of facial image data and the design, training and evaluation of the proposed convolutional neural network (CNN) models for facial affect analysis and recognition are described in Sec.~\ref{sec:aff}. A detailed description of the application implementation, along with illustrations, is provided in Sec.~\ref{sec:app}. Conclusions and discussion together with suggestions for future work are presented in Sec.~\ref{sec:con}. 

The source code for the application and the machine learning setup is available on GitHub~\cite{emosic}.

\subsection{Related Work}

Commercially available tools for real world affect analysis and recognition are fairly limited. Affectiva~\cite{affectiva} is the most established company, offering a number of successful applications, for example in adaptive children's computer games, automatic tagging on the Imgur image hosting site and assessing viewer reception of television adverts. Microsoft are also trialling Emotion API~\cite{microsoftemotion} which offers similar functionality, though has so far seen little in terms of real world applications.

Small-scale deployments of automatic affect recognition have generally focussed on video games~\cite{videogames}, medical applications~\cite{chronicpain, autism} and analysis of driver emotions~\cite{driverfrustration}. There has so far been very little development of EIUIs. This may in part be due to user reluctance based on privacy concerns~\cite{reynolds2005evaluation}, as well as technological limitations.

To date, mobile affective computing has mostly remained limited to activity monitoring based on accelerometer data and calls, SMS and application usage~\cite{mobAC2014, surveymobileaff} with only two examples involving input from the device camera~\cite{surveymobileaff}. Collecting and analysing visual data is generally considered a high-computational task with the need for wide deployment of cameras. However, this trend is bound to change with the availability of new hardware, datasets and machine learning approaches.

The recently released AffectNet dataset~\cite{affectnet} is a very large (450,000 images), in-the-wild annotated dataset for training automatic affect recognition models. The dataset provides annotations for 8 emotion categories, valence and arousal on a continuous scale from -1 to 1, facial bounding boxes and 66 facial landmarks. Previous in-the-wild datasets were generally smaller and did not include annotations of valence and arousal. For instance, FER-2013~\cite{fer2013} included 35,000 images with 7 emotion categories, FER-Wild~\cite{ferwild} included 25,000 similarly annotated images and EmotioNet~\cite{emotionet} contained 100,000 images with 23 emotion categories. The increased availability of these large annotated datasets of facial images enables further developments in the field of affective computing.

Existing machine learning approaches typically do not consider model size as an important attribute in architecture design. General image classification architectures such as InceptionV3~\cite{inceptionnet}, ResNet50~\cite{resnet}  and VGG16~\cite{vggnet} result in data files of significant size; 90MB, 97MB and 528MB respectively. The CNNEmotions architecture~\cite{cnnemotion} designed for the task of emotion classification is certainly too large for any realistic mobile application (475MB), as is VGGFace~\cite{vggface}. The only architecture specifically designed with mobile deployment in mind is MobileNet~\cite{mobilenet} which, at 16.4MB is certainly reasonable for mobile deployment\footnote{All sizes relate to pre-trained CoreML~\cite{coreml} models available from CoreML Store~\cite{coremlstore}.}.

Implementations of light-weight CNN architectures for affect analysis and recognition have focussed primarily on real-time classification, and therefore often produce smaller models as a side effect of this.~\cite{realtime},~\cite{realtime2} and~\cite{happynet} specify models that achieve quite high classification accuracy ($\sim$60\% on FER2013) for frames in video feeds in real time, with file sizes generally smaller than 30MB\@.

\section{Facial Affect Recognition}\label{sec:aff}

\subsection{Considerations and Design}
Given the goal of mobile deployment, the final model size must remain reasonable for inclusion in a mobile application. Google and Apple both impose limits on app size in the Play Store and AppStore respectively. For installation over cellular network, Apple limits apps to 150MB (100MB before Sep 2017) and for all apps Google imposes a limit of 100MB (50MB before Sep 2015). Most affective computing models included in a user application are likely to augment interaction (e.g., EIUI) rather than providing the primary functionality of the app. The storage space used by these models should therefore remain well under this 100MB limit.

Cloud offload might be seen as the obvious solution to the issue of constrained resources, but for the proposed work users' facial imagery is unavoidably involved, so privacy becomes an immediate concern. Due to both privacy concerns and concerns regarding latency, local execution is the preferred course of action. 

In order to emulate a more complex application with multiple models we include two separate models---one for emotion classification and one for valence/arousal prediction---despite the clear possibility to exploit the similarity of these tasks and use a single network with multiple outputs for this application. Consequently, we impose a maximum model size of 15MB for the proposed application. The total contributed file-size should therefore remain less than 30MB for the two models, making the app approximately 50MB in size overall.

The time and computational resource taken to obtain predictions from images are also a factor to consider on mobile devices. However, this is expected to be of little issue for the models designed and implemented in this work, given the simplicity inherent in architectures of this size.

Inspired by previously established networks, three CNN architectures are designed and evaluated: (1) a design similar to AlexNet~\cite{alexnet} using a series of convolution layers with incrementally smaller kernels interspersed with max-pooling layers, (2) an architecture based on VGG16~\cite{vggnet} with stacked $3\times3$ convolution layers interspersed with max-pooling layers, and (3) a network based on MobileNet~\cite{mobilenet} utilising depth-wise separable convolutions to maximise spatial efficiency.

All CNN models are implemented using Keras~\cite{keras} and trained on an NVIDIA GeFore GTX 1080 Ti GPU using TensorFlow~\cite{tensorflow}.

\subsection{Preprocessing and Training}\label{sec:aff:pre}

The AffectNet dataset~\cite{affectnet} contains images of a highly heterogeneous nature. The dataset is divided by its creators into training and validation sets, and the test set labels are not yet available for research purposes. To be able to compare our results to that of the baseline as reported in~\cite{affectnet}, we follow the predefined dataset partitions. 

In order to produce suitable images for input to a CNN the faces are cropped and resized to $128\times128$ pixels. The facial bounding box annotations provided by AffectNet are used for this purpose. Only manually annotated images are used\footnote{AffectNet also includes a large number of images automatically annotated by models trained on the manually annotated images.}. For emotion classification all images annotated with invalid emotions (8: none, 9: uncertain and 10: no-face) are discarded leaving a total training set of 287,651 images and a validation set of 4000 images. For valence/arousal regression all images with invalid annotations, indicated using a value of -2, are discarded leaving a training set of 320,739 images and a validation set of 4500 images.

Weighted-loss is used for emotion classification to account for the imbalance in the training set as this achieved the best results in the baseline paper~\cite{affectnet} (compared with up- and down-sampling). For valence/arousal regression data imbalance is again a problem resulting in over-fitting and potentially reduced performance. The mean annotations of the training set are 0.19 and 0.09 for valence and arousal respectively, while for the validation set are -0.16 and 0.30. Attempting to rectify this by down-sampling did little to improve performance so the full training set is used.

Randomised data augmentation is used for the training set with potential for images to be rotated by up to 20 degrees, translated by up to 10\% (in both $x$- and $y$-directions) and flipped in the $x$-direction. All image data is normalised from $[0, 255]$ to $[0, 1]$ to increase the speed of training.

The Adam optimiser~\cite{adamopt} is used throughout with suggested parameters $\alpha=0.001$, $\beta_1=0.9$, $\beta_2=0.999$ and $\epsilon=10^{-8}$, this is due to its design focus for machine learning tasks on large datasets. Batch size is maximised in order to best encapsulate the varied nature of the data and therefore improve training; 400 for architectures 1 and 2, and 250 for architecture 3, limited by available memory on the training hardware.

All classification models are trained over 24 epochs. As there is a strong correlation between valence/arousal and emotion, transfer learning can be exploited to produce the required valence/arousal models more easily. As such, the output layers of the trained emotion classifiers can be removed and replaced with appropriate output layers for the regression task (described below). 
The resulting models are then fine-tuned over 16 epochs. Both training times were chosen based on the details provided in the AffectNet baseline paper~\cite{affectnet} and resulted in a plateau in validation loss towards the end of training.

\subsection{Architecture 1: AlexNet Variant}
This architecture is inspired by AlexNet~\cite{alexnet}, including a series of incrementally smaller convolution kernels starting at $9\times9$ and reducing to $3\times3$ with $2\times2$ max-pooling layers in between each convolution block and two fully connected (dense) layers prior to the output layer. There is a 0.2 Gaussian dropout after each pooling layer and a 0.5 dropout after each dense layer. Unlike the AlexNet architecture, each convolution block is constructed from a conventional 2D convolution layer followed by a batch normalisation layer~\cite{batchnorm} and a ReLU activation layer~\cite{relu}. This helps to provide regularisation and faster training. The architecture is also shallower and narrower than the original AlexNet design in order to minimise model size. The full architecture specification is given in Table~\ref{tab:arch:alex}; the output layer contains 8 nodes with soft-max activation for emotion classification and 2 nodes with linear activation for valence/arousal regression.

\subsection{Architecture 2: VGGNet Variant}
This architecture is fairly similar to the AlexNet inspired design above, though it uses the principle behind VGG16~\cite{vggnet} of stacked $3\times3$ convolution kernels to capture larger image structure. The convolution blocks described for Arch. 1 above are again used, interspersed with max-pooling layers and followed by two fully connected layers before the output layer. As above, each pooling layer is followed by a 0.2 Gaussian dropout and there is a 0.5 dropout after each dense layer. The full architecture is given in Table~\ref{tab:arch:vgg}, it is also narrower and shallower than typical VGGNet implementations in order to conserve space.

\begin{table}
    \centering
    \caption{CNN architecture 1: AlexNet variant.\label{tab:arch:alex}}
    \footnotesize
    \begin{tabular}{@{}ccc@{}}
        \toprule
        Type            & Shape               & Output                   \\ \midrule
        Conv            & $9\times9\times16$  & $128\times128\times16$   \\
        MaxPool         & $2\times2$          & $64\times64\times16$     \\
        Conv            & $7\times7\times32$  & $64\times64\times32$     \\
        MaxPool         & $2\times2$          & $32\times32\times32$     \\
        Conv            & $5\times5\times64$  & $32\times32\times64$     \\
        MaxPool         & $2\times2$          & $16\times16\times64$     \\
        Conv            & $3\times3\times128$ & $16\times16\times128$    \\
        MaxPool         & $2\times2$          & $8\times8\times128$      \\
        Conv            & $3\times3\times128$ & $8\times8\times128$      \\
        MaxPool         & $2\times2$          & $4\times4\times128$      \\
        Flatten         & 2048                & $-$                      \\
        $2\times$Dense  & 1024                & $-$                      \\
        Dense           & 8 or 2              & 1 label or 2 floats      \\
        \bottomrule
    \end{tabular}
\end{table}

\begin{table}
    \centering
    \caption{CNN architecture 2: VGGNet variant.\label{tab:arch:vgg}}
    \footnotesize
    \begin{tabular}{@{}ccc@{}}
        \toprule
        Type            & Shape               & Output                  \\ \midrule
        $2\times$Conv   & $3\times3\times16$  & $128\times128\times16$  \\
        MaxPool         & $2\times2$          & $64\times64\times16$    \\
        $2\times$Conv   & $3\times3\times32$  & $64\times64\times32$    \\
        MaxPool         & $2\times2$          & $32\times32\times32$    \\
        $2\times$Conv   & $3\times3\times64$  & $32\times32\times64$    \\
        MaxPool         & $2\times2$          & $16\times16\times64$    \\
        $2\times$Conv   & $3\times3\times128$ & $16\times16\times128$   \\
        MaxPool         & $2\times2$          & $8\times8\times128$     \\
        $2\times$Conv   & $3\times3\times128$ & $8\times8\times128$     \\
        MaxPool         & $2\times2$          & $4\times4\times128$     \\
        Flatten         & 2048                & $-$                     \\
        $2\times$Dense  & 1024                & $-$                     \\
        Dense           & 8 or 2              & 1 label or 2 floats     \\
        \bottomrule
    \end{tabular}
\end{table}

\begin{table}
    \centering
    \caption{CNN architecture 3: MobileNet variant.\label{tab:arch:mob}}
    \footnotesize
    \begin{tabular}{@{}cccc@{}}
        \toprule
        Type            & Shape                 & Stride        & Output                 \\ \midrule
        Conv            & $3\times3\times32$    & 2             & $64\times64\times32$   \\
        DConv           & $3\times3\times64$    & 1             & $64\times64\times64$   \\
        DConv           & $3\times3\times128$   & 2             & $32\times32\times128$  \\
        DConv           & $3\times3\times128$   & 1             & $32\times32\times128$  \\
        DConv           & $3\times3\times256$   & 2             & $16\times16\times256$  \\
        DConv           & $3\times3\times256$   & 1             & $16\times16\times256$  \\
        DConv           & $3\times3\times512$   & 2             & $8\times8\times512$    \\
        $5 \times$DConv & $3\times3\times512$   & 1             & $8\times8\times512$    \\
        DConv           & $3\times3\times1024$  & 2             & $4\times4\times1024$   \\
        DConv           & $3\times3\times1024$  & 1             & $4\times4\times1024$   \\
        GlobalAvePool   & 1024                  & $-$           & $-$                    \\
        Dense           & 8 or 2                & $-$           & 1 label or 2 floats    \\
        \bottomrule
    \end{tabular}
\end{table}

\subsection{Architecture 3: MobileNet Variant}
This architecture is inspired by MobileNet~\cite{mobilenet}, which leverages $3\times3$ depth-wise separable convolution layers followed by $1\times1$ conventional convolution layers to retain high performance while minimising architectural complexity. This results in far smaller, tunable, network architectures perfect for deployment to mobile devices. Depth-wise separable convolution (DConv) blocks as described in~\cite{mobilenet} are used, with the full architecture given in Table~\ref{tab:arch:mob}. The reduced layer-wise complexity allows for a much deeper model which also retains good width. The output layer remains as above, but no pooling layers are present (stride in convolution layers is instead used for down-sampling) other than the final global average pooling layer which replaces the conventional fully connected layers. This pooling layer is followed by a dropout at rate 0.3.

\subsection{Evaluation Results}\label{sec:aff:res}

All architectures are evaluated on the AffectNet validation set (the test set is not publicly available) using the metrics provided for the baselines in~\cite{affectnet}. Human annotator agreement for emotion classification on AffectNet is just over 60\%.

For emotion classification accuracy (ACC), F1-score (F1), Cohen's kappa~\cite{cohen} (KAPPA), Krippendorff's alpha~\cite{kalpha} (ALPHA), area under precision-recall curve (AUCPR) and area under ROC curve (AUC) are used. For valence/arousal prediction RMSE, Pearson's correlation coefficient (CORR), sign agreement metric~\cite{sagr} (SAGR) and concordance correlation coefficient~\cite{concorcoe} (CCC) are used.

Emotion classification results are presented in Table~\ref{tab:class_arch_results}. The table shows that the VGGNet variant outperforms the AlexNet variant and the MobileNet variant in all metrics. It also outperforms the baseline in all but accuracy and F1 which are equalled at 58\%.

\begin{table}[b]
    \centering
    \caption{Emotion classification performance metrics for each architecture against weighted-loss baseline.\label{tab:class_arch_results}}
    \begin{tabular}{@{}lrrrr@{}}
        \toprule
                                      & Baseline         & Arch. 1       & Arch. 2       & Arch. 3 \\ \midrule
        \multicolumn{1}{@{}l}{ACC}    & \textbf{0.58}    & 0.56          & \textbf{0.58} & 0.56    \\
        \multicolumn{1}{@{}l}{F1}     & \textbf{0.58}    & 0.56          & \textbf{0.58} & 0.56    \\
        \multicolumn{1}{@{}l}{KAPPPA} & 0.51             & 0.50          & \textbf{0.52} & 0.50    \\
        \multicolumn{1}{@{}l}{ALPHA}  & 0.51             & 0.50          & \textbf{0.52} & 0.50    \\
        \multicolumn{1}{@{}l}{AUCPR}  & 0.56             & 0.61          & \textbf{0.62} & 0.60    \\
        \multicolumn{1}{@{}l}{AUC}    & 0.82             & \textbf{0.90} & \textbf{0.90} & 0.89    \\
        \bottomrule
    \end{tabular}
\end{table}

Valence/arousal regression results are shown in Table~\ref{tab:regr_arch_results}. As with emotion classification, the VGGNet variant provides the best results of the three proposed architectures, for both valence and arousal prediction, though only marginally. All proposed architectures perform better for arousal than for valence, also outperforming the baseline. In contrast, the baseline performs significantly better for valence than arousal, also outperforming all proposed architectures.

\begin{table}
\centering
\caption{Valence (V) and arousal (A) regression performance metrics for each architecture against weighted-loss baseline.\label{tab:regr_arch_results}}
\resizebox{\columnwidth}{!}{%
\begin{tabular}{@{}lccccccccccc@{}}
\toprule
& \multicolumn{2}{c}{Baseline}  && \multicolumn{2}{c}{Arch. 1}  && \multicolumn{2}{c}{Arch. 2}    && \multicolumn{2}{c}{Arch. 3}  \\
\cmidrule{2-3} \cmidrule{5-6} \cmidrule{8-9} \cmidrule{11-12}
                            & V             & A             && V             & A            && V             & A              && V             & A            \\ \midrule
\multicolumn{1}{@{}l}{RMSE} & \textbf{0.37} & 0.41          && 0.41          & 0.39         && 0.41          & \textbf{0.37}  && 0.42          & 0.38         \\
\multicolumn{1}{@{}l}{CORR} & \textbf{0.66} & 0.54          && 0.59          & 0.53         && 0.62          & \textbf{0.56}  && 0.59          & 0.53         \\
\multicolumn{1}{@{}l}{SAGR} & 0.74          & 0.65          && 0.73          & 0.74         && \textbf{0.75} & \textbf{0.75}  && 0.73          & 0.74         \\
\multicolumn{1}{@{}l}{CCC}  & \textbf{0.60} & 0.34          && 0.54          & 0.43         && 0.57          & \textbf{0.48}  && 0.55          & 0.47         \\
\bottomrule
\end{tabular}%
}
\end{table}

\subsection{Analyses and Discussion}\label{sec:aff:analy}

The increased spatial efficiency of the MobileNet variant, and consequently its greater depth and width, do surprisingly little to improve the performance of the model over the AlexNet and VGGNet variants. There are many potential reasons for these results. One possible explanation is that facial affect might rely on edge related features which are typically captured by max pooling, but MobileNets only use average pooling. Another potential reason is the slight variation in model size (the VGGNet variant is slightly bigger), or the increased use of dropout in VGGNet.

All models have a file size close to the goal of 15MB, with the VGGNet variant being the largest at 15MB and the MobileNet variant the smallest at 13.2MB\@. All of these remain viable for mobile deployment as described in the considerations, and have performance close to the baseline for the AffectNet dataset.

Table~\ref{tab:conf-mat} shows the confusion matrix of the VGGNet variant, the best performing architecture, providing a classification breakdown for \textbf{N}eutral, \textbf{H}appy, \textbf{Sa}d, \textbf{Su}prised, \textbf{Af}raid, \textbf{D}isgusted, \textbf{An}gry and \textbf{C}ontemptuous for the validation set containing 500 examples of each emotion. We observe that \textit{happiness} has the highest rate of correct classifications (72\%), while \textit{anger} has the lowest with just 43\% correct, often being confused with \textit{disgust}. In the literature, anger and disgust are known to be confused because of the facial action units they share~\cite{wiggers82}. 

\begin{table}[]
    \centering
    \caption{Emotion classification confusion matrix of the VGGNet variant for the AffectNet validation set.\label{tab:conf-mat}}
    \begin{tabular}{lEEEEEEEE}
        \multicolumn{1}{c|}{}  & \multicolumn{1}{c}{N} & \multicolumn{1}{c}{H}   & \multicolumn{1}{c}{Sa}   & \multicolumn{1}{c}{Su}  & \multicolumn{1}{c}{Af}  & \multicolumn{1}{c}{D}   & \multicolumn{1}{c}{An}   & \multicolumn{1}{c}{C}   \\ \hline
        \multicolumn{1}{l|}{N}  & 247 & 7   & 52  & 60  & 11  & 22  & 34  & 67  \\
        \multicolumn{1}{l|}{H}  & 20  & 358 & 6   & 26  & 4   & 15  & 4   & 67  \\
        \multicolumn{1}{l|}{Sa} & 63  & 9   & 279 & 22  & 38  & 41  & 37  & 11  \\
        \multicolumn{1}{l|}{Su} & 33  & 22  & 15  & 298 & 97  & 20  & 7   & 8   \\
        \multicolumn{1}{l|}{Af} & 21  & 6   & 32  & 72  & 320 & 32  & 12  & 5   \\
        \multicolumn{1}{l|}{D}  & 29  & 9   & 36  & 24  & 31  & 316 & 42  & 13  \\
        \multicolumn{1}{l|}{An} & 71  & 4   & 39  & 22  & 29  & 98  & 216 & 21  \\
        \multicolumn{1}{l|}{C}  & 71  & 56  & 12  & 21  & 3   & 33  & 26  & 278 \\
    \end{tabular}
\end{table}

\section{Mobile Music Recommendation}\label{sec:app}

\subsection{User Interface}
\begin{figure}
    \centering
    \begin{subfigure}{0.48\linewidth}
        \includegraphics[width=\linewidth]{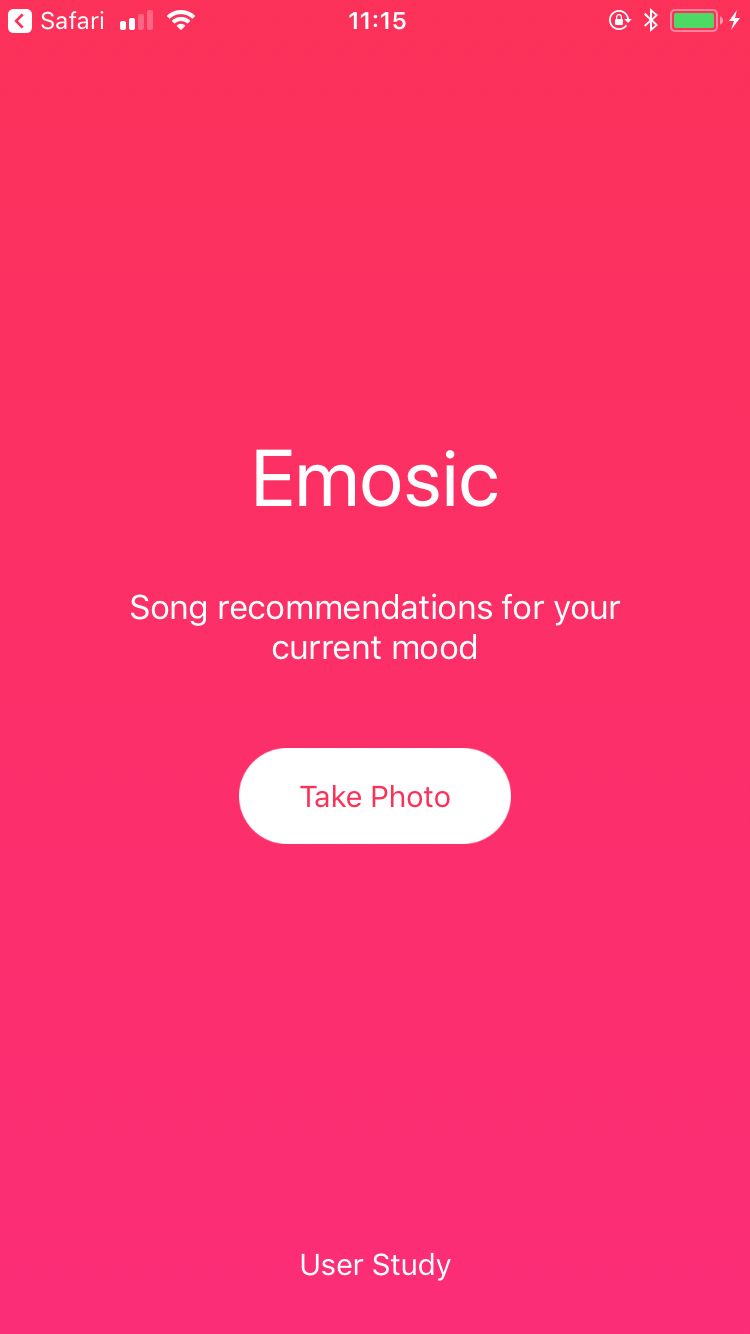}
        \subcaption{Main screen.\label{fig:mainui:front}}
	\end{subfigure}
    \hfill
    \begin{subfigure}{0.48\linewidth}
       \includegraphics[width=\linewidth]{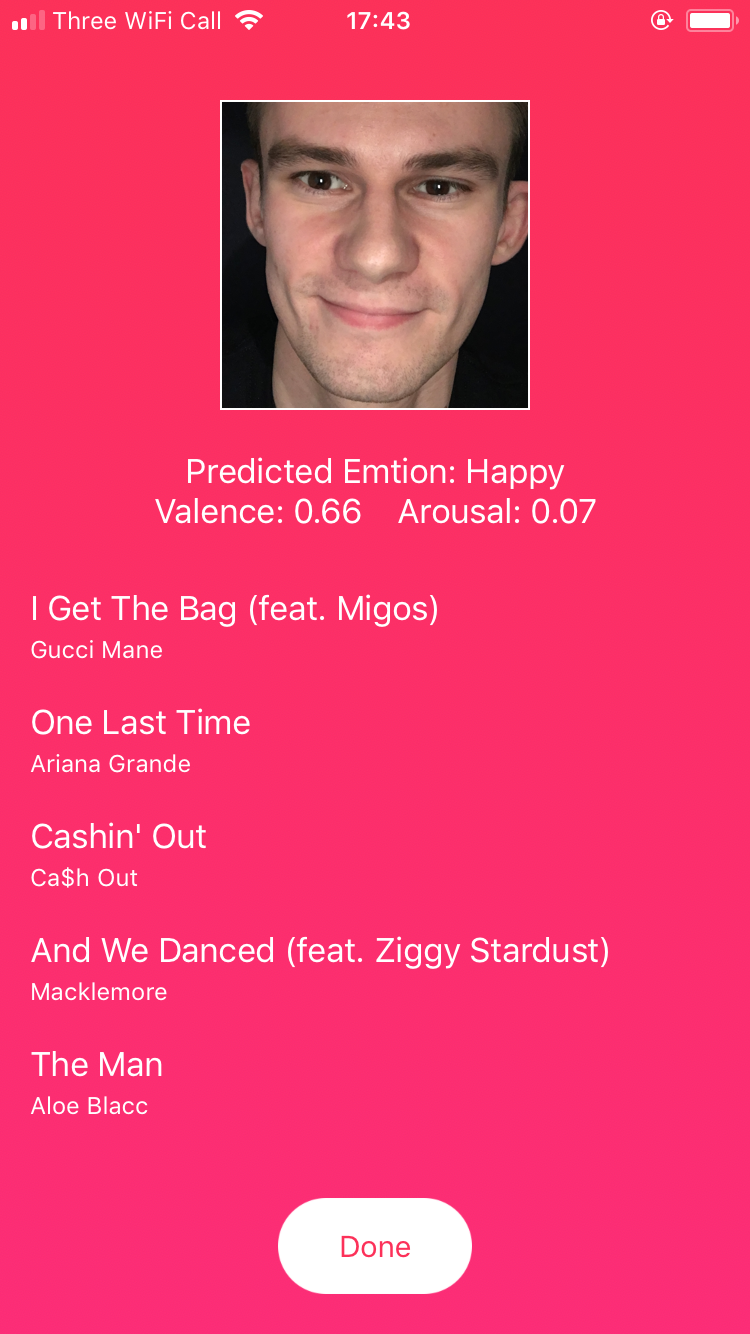}
       \subcaption{Song recommendation.\label{fig:mainui:rec}}
    \end{subfigure}
    \caption{Emosic app user interface.\label{fig:mainui}}
\end{figure}

The Emosic mobile application is implemented in Swift for the iOS platform and has a very simple interface as shown in Fig.~\ref{fig:mainui}. The user opts to take a photo, which prompts the native camera interface to be presented using the front-facing camera. Once the user has taken a photo of their face, the emotion, valence and arousal are predicted and the results are presented on the screen in Fig.~\ref{fig:mainui:rec}. In normal operation, the predicted emotion, valence and arousal are shown to the user along with the top five recommended songs. Clicking on each song opens it in the Spotify app~\cite{spotifyapp} for the user to listen to.

\subsection{Facial Affect Analysis}
In order to be used for prediction within the iOS app, the highest performing Keras models described in Sec.~\ref{sec:aff} (i.e., Arch.~2) are converted for use with Apple's CoreML framework~\cite{coreml}. Apple provides an open source Python tool-kit, Coremltools~\cite{coremltools}, for this purpose. Almost no additional modification is required for the models to function within the iOS app, only minor preprocessing of the input image data. To match the input format described in Sec.~\ref{sec:aff:pre}, Apple's Vision framework~\cite{visionapi} is used to determine the bounding box for the user's face, this is then cropped and resized to the required $128\times128$ pixels. The image data can then be fed directly to the applicable model after conversion to pixel buffer format.

\subsection{Song Recommendation}\label{sec:app:rec}
Song recommendations are obtained using the Spotify Web API recommendations service~\cite{spotify}. This service provides a REST endpoint which can be given up to five seed genres, a modality (major or minor) and numeric values for valence and energy (taken to be analogous to arousal) between 0 and 1. A list of songs which best match the inputs are then returned in JSON format. Seed genres are determined using a predefined mapping from the emotion predicted by the CoreML classifier (one of the basic eight) to a list of five seed genres. Predicted valence and arousal from the CoreML regression model are used directly after translation from the output range of $[-1,1]$ to the required $[0,1]$. The desired modality is taken from the sign of the valence prediction, positive being major and negative minor.

\subsection{User Study}
\begin{figure*}
    \centering
    \subcaptionbox{Instructing the user which emotion to emulate.\label{fig:usui:inst}}[0.28\linewidth][c]{%
        \includegraphics[width=0.24\linewidth]{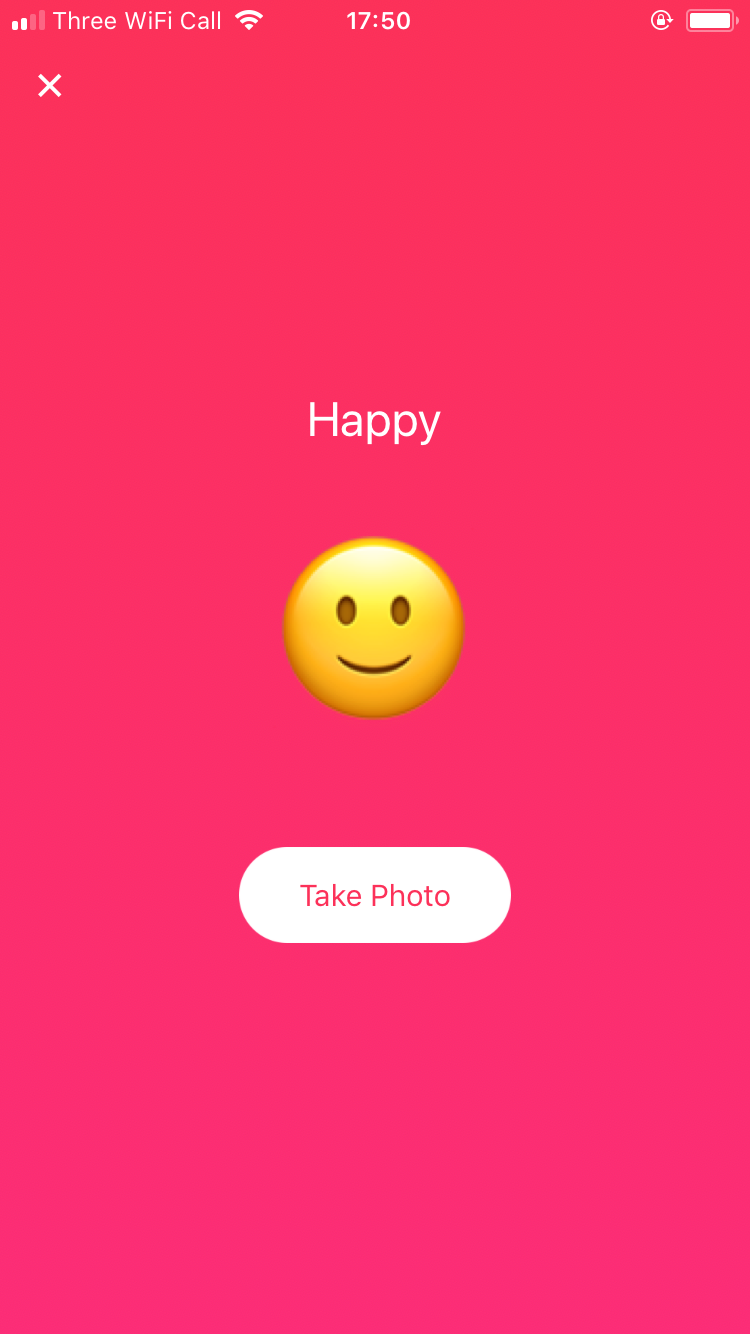}%
	}
    \hfill
    \subcaptionbox{Song recommendation with rating input.\label{fig:usui:rec}}[0.28\linewidth][c]{%
        \includegraphics[width=0.24\linewidth]{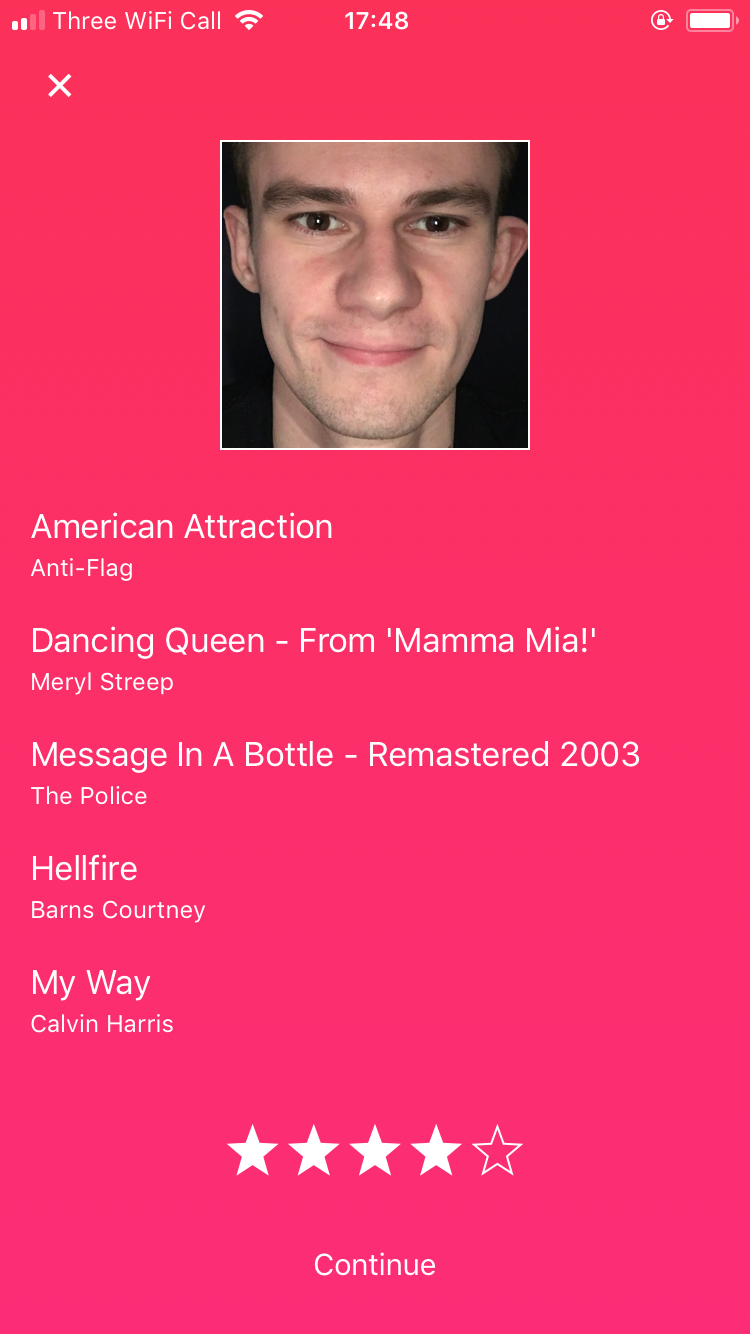}%
	}
    \hfill
    \subcaptionbox{Self annotation of displayed facial affect with valence and arousal sliders.\label{fig:usui:ann}}[0.28\linewidth][c]{%
        \includegraphics[width=0.24\linewidth]{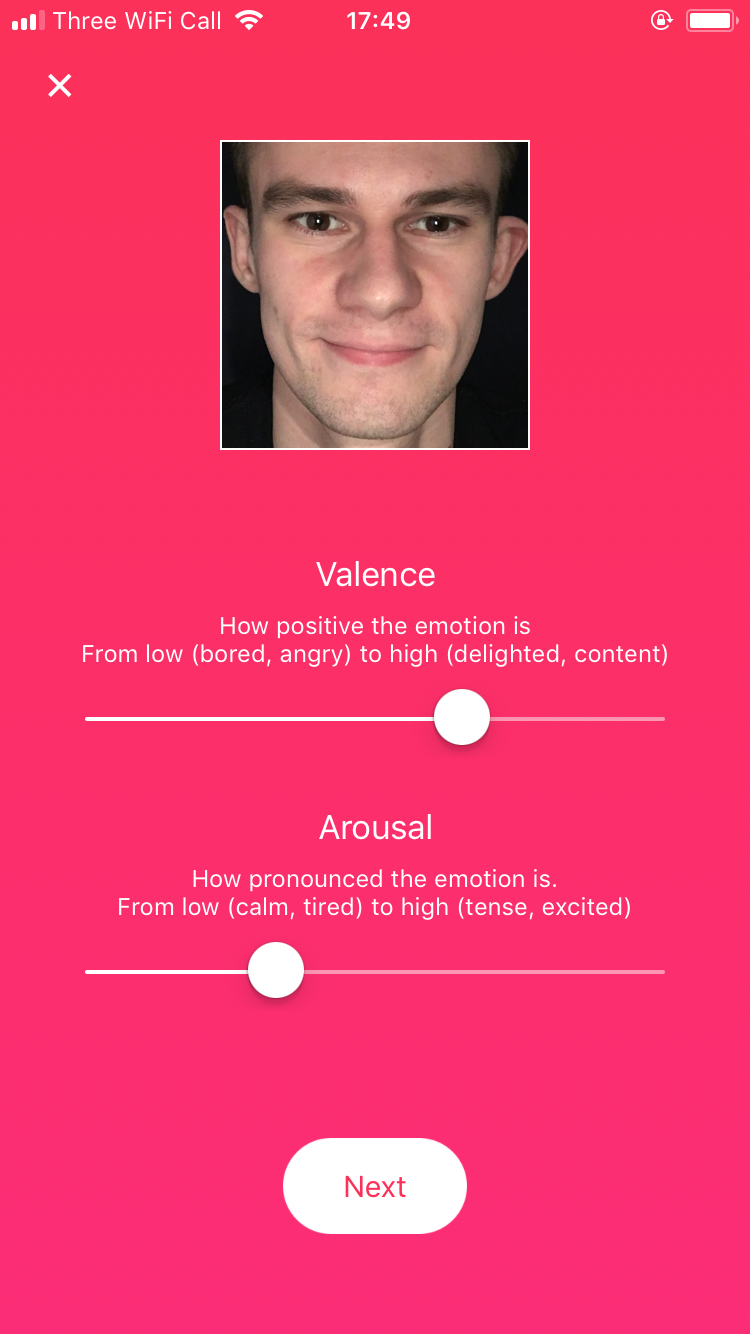}%
	}
    \caption{User study interface.\label{fig:usui}}
\end{figure*}

The user study is built directly into the app and can be accessed from the bottom of the screen as shown in Fig.~\ref{fig:mainui:front}. The user is first given instructions regarding the study and information about data retention, they are then presented with the screens shown in Fig.~\ref{fig:usui} in order from left to right.

Firstly, the user is presented an emotion  which they are to emulate (Fig~\ref{fig:usui:inst}). Then they take a photo using the native camera interface. The recommended songs are presented with a 5-star rating input (Fig.~\ref{fig:usui:rec}). Subsequently, a self-annotation screen for valence and arousal (Fig.~\ref{fig:usui:ann}) is displayed. This sequence of screens is displayed for ten distinct emotions: neutral, delighted, happy, miserable, sad, surprised, angry, afraid, disgusted and contemptuous. These emotion categories are chosen (i) to correspond closely with the eight emotions of AffectNet that the models are trained with, and (ii) to provide some notable variance along valence and arousal dimensions.

The study was completed by 10 participants (4 female, 6 male) aged between 19 and 54 (mean 27, std 14). The results are broken down as in Sec.~\ref{sec:aff:res}, with recognised emotion classes compared against instructed emotions, and valence/arousal predictions compared against self-annotated values. Many participants reported that the valence/arousal annotation scheme was not intuitive, this has likely led to discrepancies in what is taken to be ground-truth for the study, particularly for arousal.

The predictions made by the deployed models in the user study are evaluated using the same metrics as for the AffectNet dataset described in Sec.~\ref{sec:aff:res}. The recognition and classification results are shown in Tables~\ref{tab:us_res:a} and~\ref{tab:us_res:b}. The confusion matrix of emotion classification for the user study is given in Table~\ref{tab:us_conf_mat}. When measuring classification performance, in order to obtain a balanced set which is directly comparable to the results in Sec.~\ref{sec:aff:res}, the results for \textit{happy} and \textit{sad} are represented using the emotions \textit{delighted} and \textit{miserable}.

\begin{table}
    \centering
    \caption{User study results.\label{tab:us_res}}
    \begin{subfigure}[t]{0.4\linewidth}
        \centering
        \caption{Emotion classification.\label{tab:us_res:a}}
        \begin{tabular}{@{}lr@{}}
            \toprule
            \multicolumn{1}{@{}l}{ACC}    & 0.49  \\
            \multicolumn{1}{@{}l}{F1}     & 0.45  \\
            \multicolumn{1}{@{}l}{KAPPPA} & 0.41  \\
            \multicolumn{1}{@{}l}{ALPHA}  & 0.41  \\
            \multicolumn{1}{@{}l}{AUCPR}  & 0.60  \\
            \multicolumn{1}{@{}l}{AUC}    & 0.88  \\
            \bottomrule
        \end{tabular}
    \end{subfigure}
    \hfill
    \begin{subfigure}[t]{0.55\linewidth}
        \centering
        \caption{Valence/Arousal prediction.\label{tab:us_res:b}}
        \begin{tabular}{@{}lrr@{}}
            \toprule
                                        & Valence       & Arousal \\ \midrule
            \multicolumn{1}{@{}l}{RMSE} & 0.40          & 0.40    \\
            \multicolumn{1}{@{}l}{CORR} & 0.74          & 0.44    \\
            \multicolumn{1}{@{}l}{SAGR} & 0.74          & 0.65    \\
            \multicolumn{1}{@{}l}{CCC}  & 0.68          & 0.39    \\
            \bottomrule
        \end{tabular}
    \end{subfigure}
\end{table}

\begin{table}
    \centering
    \caption{User study emotion classification confusion matrix.\label{tab:us_conf_mat}}
    \begin{tabular}{lFFFFFFFF}
        \multicolumn{1}{c|}{}   & \multicolumn{1}{c}{N} & \multicolumn{1}{c}{H}   & \multicolumn{1}{c}{Sa}   & \multicolumn{1}{c}{Su}  & \multicolumn{1}{c}{Af}  & \multicolumn{1}{c}{D}   & \multicolumn{1}{c}{An}   & \multicolumn{1}{c}{C}   \\ \hline
        \multicolumn{1}{l|}{N}  & 3  & 1  & 0   & 3   & 0   & 0   & 1   & 2   \\
        \multicolumn{1}{l|}{H}  & 0  & 9  & 0   & 0   & 1   & 0   & 0   & 0   \\
        \multicolumn{1}{l|}{Sa} & 2  & 0  & 7   & 0   & 0   & 0   & 0   & 1   \\
        \multicolumn{1}{l|}{Su} & 0  & 0  & 0   & 4   & 6   & 0   & 0   & 0   \\
        \multicolumn{1}{l|}{Af} & 1  & 0  & 1   & 1   & 7   & 0   & 0   & 0   \\
        \multicolumn{1}{l|}{D}  & 0  & 1  & 1   & 0   & 1   & 7   & 0   & 0   \\
        \multicolumn{1}{l|}{An} & 1  & 0  & 2   & 0   & 2   & 3   & 1   & 1   \\
        \multicolumn{1}{l|}{C}  & 2  & 1  & 2   & 1   & 1   & 2   & 0   & 1   \\
    \end{tabular}
\end{table}

\subsection{Analyses and Discussion}

Emotion classification results are slightly worse than for the AffectNet validation set and vary greatly between emotions. \textit{Happy} has an accuracy of 90\% while \textit{contempt} is at just 10\%. \textit{Surprise} is often misclassified as \textit{fear}, and \textit{anger} as \textit{disgust}, similar to the results for AffectNet in Sec.~\ref{sec:aff:res}. Participants reported that some emotions were difficult to emulate in this context (e.g.\ \textit{contempt}) which may have caused variation in what is deemed to be ground-truth. Valence/arousal prediction is more successful, broadly matching the AffectNet results. Valence has a notably improved CORR and CCC scores of 0.74 and 0.68, while arousal has generally worse performance, more closely matching the AffectNet baseline.

Average runtime of the deployed CoreML models on an iPhone 6S was 22.4ms across ten runs. This equates to approximately 45 fps, suggesting that the models are well suited for real-time deployment on video streams.

The average rating for song recommendations was 3.17, indicating that users had a generally positive view of the application's functionality. Emotion specific ratings are shown in Table~\ref{tab:us_ratings}. Participants reported that emotions with clearer connotations (e.g.\ \textit{happy}, \textit{sad}) were easier to interpret musically and therefore easier to rate, while emotions with less clear connotations (e.g.\ \textit{contempt}, \textit{disgust}) were reported as being difficult to interpret.
The lower ratings for the latter may be due to users finding them harder to relate to any music, or these emotions being classified correctly less often and therefore ending up with inappropriate music recommendation. The fact that the user rating was slightly higher (3.31) when emotion was classified correctly than when the predicted emotion was incorrect (3.02) supports the second of these suggestions.

\begin{table}
    \centering
    \caption{User study song recommendation ratings.\label{tab:us_ratings}}
    \begin{tabular}{@{}lr@{}}
            \toprule
            \multicolumn{1}{@{}l}{Neutral}      & 3.6          \\
            \multicolumn{1}{@{}l}{Delighted}    & 3.5          \\
            \multicolumn{1}{@{}l}{Happy}        & 3.9          \\
            \multicolumn{1}{@{}l}{Miserable}    & 3.2          \\
            \multicolumn{1}{@{}l}{Sad}          & 3.7          \\
            \multicolumn{1}{@{}l}{Surprised}    & 3.2          \\
            \multicolumn{1}{@{}l}{Angry}        & 2.7          \\
            \multicolumn{1}{@{}l}{Afraid}       & 2.4          \\
            \multicolumn{1}{@{}l}{Disgusted}    & 2.6          \\
            \multicolumn{1}{@{}l}{Contemptuous} & 2.9          \\ \midrule
            \multicolumn{1}{@{}l}{Average}        & 3.2          \\
            \bottomrule
        \end{tabular}
\end{table}

A possible solution to the above mentioned issue is using a music-specific emotion model for labelling and assessment. One such model is derived from the Geneva Emotion Music Scale (GEMS) and has been developed for musically induced emotions~\cite{ZentnerEtAl2008}. It consists of nine emotional scales: wonder, transcendence, tenderness, nostalgia, peacefulness, power, joyful activation, tension and sadness. Zentner et al.~\cite{ZentnerEtAl2008} compared the discrete approach, the dimensional approach and the GEMS approach and reported that participants preferred to report their emotions using the GEMS approach. Therefore, future studies focusing on affect-based music recommendation should take this into consideration. However, this would require a large dataset acquired in this specific context (which is not yet available).  

During the user study, participants were also asked about their privacy concerns with the application. Most responded that they would not like the photos to be saved, and would prefer that data remained local to the device. Some also mentioned that they would like the uses for the data to be clearly stated and agreed to, though overall level of concern seemed to be significantly less than might be expected based on previous literature~\cite{reynolds2005evaluation}.

\section{Conclusions and Future Work}\label{sec:con}
In this paper three CNN architectures for facial affect analysis have been designed and evaluated with the aim of minimising storage requirements for mobile deployment. These models gave comparable results to the current baseline when evaluated on the AffectNet dataset~\cite{affectnet}. 

The best-performing models (i.e., VGGNet variants for emotion classification and for valence/arousal prediction) were deployed in a music recommendation application with an average execution time of 22.4ms ($\sim$45fps), also suitable for real-time applications. A user study was conducted to assess their real-world performance; the results showed that the deployed models provide results that are similar to the evaluation results obtained on the AffectNet dataset. Additionally, the users reported to be generally happy with the application's functionality. These results support the proposition that EIUIs are an area of great potential within affective computing and are now becoming increasingly feasible in a real-world setting.

The functionality of the Emosic application could easily be integrated into a fully-featured music application such as Spotify~\cite{spotifyapp} and expanded to great effect. For example, determining a user's affect each time they manually choose a song would allow a model to be built up over time to provide tailored predictions for that specific user. 

However, it is important to note that a user is unlikely to be as expressive as they are prompted to be in this study, which may reduce the application's effectiveness. The recent rise of wearables might provide a solution to this~\cite{GunesHung-2016}, as a number of modalities useful for affective computing, such as heartbeat, are now more readily available within mobile applications. These could easily be incorporated into multi-modal models along with accelerometer or usage activity data to improve the accuracy and reduce invasiveness of emotion recognition in a mobile setting.

It is not difficult to see how emotionally intelligent behaviour analysis could be expanded to many other application domains, though privacy issues need to be given care and consideration. Emphasis will need to be placed on clearly explaining what such applications will be doing, and keeping computation local with as little long-term data retention as possible.

To facilitate wide-scale adoption of EIUI, continued research into very efficient (both in terms of file-size and computation) deep neural network architectures will be required. Google's MobileNets~\cite{mobilenet} are a very promising start in this respect. Alternative structures such as InceptionV3~\cite{inceptionnet} may provide a more efficient basis than the options presented in this paper and lower floating point precision could be an easy way to cut model size, though the impact on performance might be significant. It is likely that major developments will need to be driven by popular smart-phone manufacturers at an OS level, as is already beginning to happen with Apple's CoreML~\cite{coreml} and Vision frameworks~\cite{visionapi}.

\bibliographystyle{ACM-Reference-Format}
\bibliography{refs}

\end{document}